\documentclass{mem}
\usepackage{natbib}\usepackage{txfonts}\usepackage{balance}
\usepackage{graphicx}
\usepackage[a4paper,breaklinks,dvipdfm]{hyperref}
\begin{document}
\def\teff{$T\rm_{eff }$}
\def\kms{$\mathrm {km s}^{-1}$}

\title{
A short observational view of black hole X-ray binaries with \textsl{INTEGRAL}
}

   \subtitle{}

\author{
V. \,Grinberg\inst{1} 
          }

\institute{
Institut f\"ur Astronomie und Astrophysik (IAAT), Universit\"at T\"ubingen, Sand 1, 72076 T\"ubingen, Germany.
\email{grinberg@astro.uni-tuebingen.de}
}

\authorrunning{Grinberg}

\titlerunning{Black hole XRBs with \textsl{INTEGRAL}}

\abstract{
Accreting black holes are unique tools to understand the physics under extreme gravity. While black hole X-ray binaries differ vastly in mass from AGN, their accretion and ejection flows are assumed to be essentially similar.  Black hole X-ray binaries or microquasars are, however, quasars for the impatient as variability timescales scale directly with mass. State changes, i.e., strong variations in emission properties, in black hole X-ray binaries can happen within hours and whole outburst cycles within months to years. But our understanding of the drivers of such changes and the contributions of individual accretion and ejection components to the overall emission is still lacking. Here, I highlight some of the \textsl{INTEGRAL}'s unique contributions to the understanding of black hole X-ray binaries through its coverage of the energies above the spectral cutoff, its long uninterrupted monitoring observations and the measurements of hard X-ray / soft $\gamma$-ray polarization.

}
\maketitle{}

\section{Introduction}

Black hole X-ray binaries (BH-XRBs) are among the most variable sources in the X-ray sky, with the variability timescales ranging from millisecond quasiperiodic oscillations (QPOs) to months- and year-long outbursts. Low mass BH-XRBs are usually transient sources, whose outburst cycles can range from under a year in H~1743$-$322, a source that has been dormant between 1978 and 2003 \citep{Parmar_2003a} and has undergone multiple outburst since then, to cycles that are presumably longer than passed since the birth of X-ray astronomy as only one outburst has been observed for such sources. High mass BH-XRBs such as Cyg X-1 are persistent, but often highly variable in their spectral shape.

BH-XRBs show distinct emission regimes (``states''): a \textsl{hard state} during which the spectrum above $\sim$2\,keV is dominated by power law emission with a photon index $\Gamma \sim 1.7$ and an exponential cutoff at $\sim$100--400\,keV and a \textsl{soft state} when thermal emission from the accretion disk is prominent with a low contribution from a steeper power law.  The states correspond to different arrangement of accretion and ejection flows.
Through an outburst, a transient BH-XRB will follow a typical pattern of state changes, from hard to soft at higher luminosities during the outburst rise, and back from soft to hard at a lower luminosities during  the decay. Intermediate states exist and are usually short-lived and transitional \citep{Fender_2004a,Homan_2005a}. On a hardness-intensity diagram, BH-XRB will trace a typical \textsf{q}-track.

Hard and soft states also show different short-term variability behavior, with distinct shapes of the power density spectra and high rms in hard state and low rms in the soft state. However, the variability is also highly dependent on the energy and the steep power law in the soft state can be very variably when present \citep{Grinberg_2014a}. Radio emission is detected in the hard state and can be resolved in elongated structures, identified as jets. In the soft state, it is highly suppressed, pointing towards an absence of (radio emitting) jets.

In the hard state, the emission above 10\,keV can be described by an exponentially cut off power law with a folding energy of $\sim$50--500\,keV. Signatures of reflection, such as a broadened iron K$\alpha$ line at 6.4\,keV, are observed, with particularly the Compton hump contributing to the curvature above 20\,keV.  A non-thermal excess emission above the cutoff, also called ``hard tail'', is sometimes present.
The cut off power law continuum is characteristic of Comptonization of soft photons coming from the accretion disk. But the morphology of the Comptonizing hot electron  plasma is highly disputed, ranging from corona geometries that surround the disk in some way \citep[e.g.][]{Haardt_1991a,Dove_1997a} to lamppost models \citep[e.g.][]{Matt_1992a,Markoff_2005a}, where the corona could subsume the role of the jet base.  The observed hard tails could be signs of either direct jet contribution or of non-thermal components to the corona. Naturally, all models are oversimplifications of reality that could be a combination of both the above approaches, with a plasma that is likely both extended and dynamic.

The question of the accretion and ejection geometry and its variability is one of the main drivers of today's BH-XRB research. Understanding where the X-ray emission is formed is paramount to understanding the activity cycles of BH-XRBs and thus also their interaction with their environment and, by analogy, the behavior of and feedback their giant siblings, the AGN. Main open questions include the amount and kind of jet contribution to the observed hard X-ray emission, the geometry of the corona, and the disk truncation, especially in the hard state.
The simplest ansatz to answering these question and distinguishing between different continuum models is direct modelling of spectra. However, even with the best broadband X-ray data available, different models will often result in statistically similarly good fits \citep{Nowak_2011a}.

There are several ways to break this degeneracy. The straightforward one is to improve the observed spectra: this includes the progress enabled by the  3-70\,keV coverage of the \textsl{NuSTAR} mission \citep[e.g.][]{Walton_2016a}, improvements in calibration and treatment of archival data, such as \textsl{RXTE}/PCA \citep{Garcia_2014a,Garcia_2016a}, and plans for future satellites with unprecedented sensitivity in the crucial 6.4\,keV area such as \textsl{XRISM} or \textsl{Athena} \citep{Nandra_2013a}.
But different accretion and ejection geometries imply not only differences in the spectral domain, but also different long-term evolution, different spectral properties above the cut-off, different (fast) variability behavior at high energies and different polarization. \textsl{INTEGRAL} is ideally suited to address these properties with its unique combination of wide field of view, long observational campaigns, gapless broad-band coverage including above 200\,keV, fast timing and hard X-ray/soft $\gamma$-ray polarization capabilities. Its capabilities are unique among any past, current, and currently approved missions. 

In the following, I will address some of \textsl{INTEGRAL}'s contribution to studies of BH-XRBs, highlighting areas where it gives unique access to source properties and behaviors. Given \textsl{INTEGRAL}'s 16 years in orbit, a short overview like this is bound to be incomplete.
It skews towards the results from recent years; for a previous review of BH-XRBs observations with \textsl{INTEGRAL} see \citet{Del_Santo_2012a}. 


\section{INTEGRAL's unique contributions}

\subsection{Long (uninterrupted) observations}

\textsl{INTEGRAL}'s large field of view and observing strategy enable long (quasi-)uninterrupted observations that are crucial to trace the spectral parameter evolution and thus changes in the accretion geometry and allow to catch the progression of the elusive state transitions. Examples include coverage of the outburst decay of Swift~J1745$-$26 \citep{Kalemci_2014a} and the failed outburst of Swift~J174510.8$-$262411 \citep{Del_Santo_2016a} as well as two recent campaigns of new BH-XRB transients, MAXI~J1820$+$070 in spring 2018 \citep{Kuulkers_2018a} and  MAXI~J1348$-$638 in winter 2019, where \textsl{INTEGRAL} traced the source through a state transition \citep{Cangemi_2019b}.

\subsection{High energy cutoff and hard excess}

The value of the high energy cutoff is a direct signature of properties of the Comptonizing region, such as the coronal temperature. While some ansatzes exist to measure the cutoff indirectly, e.g., through its effect on the reflection features in the soft X-rays \citep{Garcia_2015a}, these approaches are model-dependent and need to be tested with direct measurements.  \textsl{INTEGRAL}'s high energy coverage is thus crucial for studies such as the analysis of Cyg X-1 by \citet{Del_Santo_2013a} addressing state evolution and corona variability.


Above the spectral cutoff, excess emission (``hard tails'') have been detected in several sources with both INTEGRAL \citep[e.g.][]{Droulans_2010a,Zdziarski_2012a,Rodriguez_2015a} and its predecessors \citep[e.g.,][]{McConnell_2000a}. Such tails are only accessible via direct measurements and
their presence signals either a direct jet contribution or a significant non-thermal contribution. The existence of the hard tails has been proven with multiple detection with different instruments. But a coherent picture of their observational properties has not emerged yet, partly due to differences in models, in definitions of what constitutes a hard excess, and in state definitions used in different analyses. At the same time the feature is intrinsically variable \citep{McConnell_2002a,Joinet_2007a}; the lack of clear correlation between the variability of the hard excess with the spectra shape below the cutoff points towards different origins for these components. 

\subsection{Fast X-ray timing}

The coverage of high energy means not only direct access to spectral properties, but also to variability properties in hard X-rays. Timing with a coded mask instrument is inherently complicated, with different data extraction algorithms prone to different systematics \citep[cf.][]{Grinberg_2011a}. \citet{Cabanac_2011a} have pushed fast X-ray timing with SPI into the high energy domain looking at the power spectra of Cyg X-1 in different spectral states up to 130\,keV; they particularly show that in the soft state, the variability may increase with energy, in agreement with trends seen later at lower energies with \textsl{RXTE} \citep{Grinberg_2014a}. \citet{Huppenkothen_2017a} have used \textsl{INTEGRAL}/ISGRI in their comprehensive analysis of the very low frequency QPO from the 2015 outburst of V404~Cyg.

\subsection{The 2015 outburst of V404~Cyg}

After $\sim$25 years of quiescence, the BH-XRB V404~Cyg went into an outburst in June 2015, first detected by \textsl{Swift} and then \textsl{MAXI} and \textsl{INTEGRAL}. The extreme flaring activity on timescales of hours, with brightest flares reaching over 40\,Crab in the hard X-rays \citep{Rodriguez_2015b,Natalucci_2015a} makes V404~Cyg a  unique testbed for accretion/ejection studies. While the lightcurves and soft spectra are remarkably complex, a careful analysis of the hard X-ray spectra has revealed a behavior similar to normal BH-XRBs \citep{Sanchez-Fernandez_2017a}, with two contributions to the hard emission \citep{Jourdain_2017a}. 

Because of the high variability, understanding where any given observation falls into the context of the long-term evolution of the source is imperative and the quasi-uninterrupted \textsl{INTEGRAL} coverage of the first $\sim$two weeks of the outburst is crucial for understanding and interpretation of other observations of the source \citep[e.g.,][]{Munoz-Darias_2016a,Motta_2017a} and further enables studies of correlations between different spectral bands and thus physical constituents of the system \citep[e.g.,][]{Maitra_2017a,Hynes_2019a}. Of special importance is the fact that high level data products were made available to the community, enabling easy access to the \textsl{INTEGRAL} results to researches not usually familiar with the mission.


\subsection{Positron annihilation signature}

The 511\,keV emission line in BH-XRBs is a long thought-after feature as it would be a smoking gun for the presence of electron-positron plasma and thus reveal a high rate of positron production in the jets of these systems. 
\textsl{INTEGRAL} and especially \textsl{INTEGRAL}/SPI is the only currently working or approved space-born mission working in the necessary energy range, but the detection in BH-XRBs remained elusive until the 2015 outburst of V404~Cyg, where \citet{Siegert_2016a} detected a feature around 511\,keV in three revolutions during which the source exhibited bright flares. This makes V404~Cyg only the third microquasar with detection of the possible positron annihilation feature, the other two being historical detections in 1E~1740.7$-$2942 and GRS~1124$-$683 with Sigma/GRANAT. The complexity of the data analysis necessary to detect the variable 511\,keV line against the high instrumental background and the variable behavior of the source prompted a critique of the detection by \citet{Roques_2016a,Roques_2019a} that the  the authors of \citet{Siegert_2016a} have addressed in further comments on data analysis included in the ArXiv version of their work (arXiv:1603.01169 v.2).

\subsection{Hard X-ray/soft $\gamma$-ray polarization}

With planned missions such as \textsl{IXPE} \citep{Weisskop_2016a} and \textsl{eXTP} \citep{Zhang_2016a} we are now entering the age of X-ray polarimetry. However, none of these forthcoming missions address energies above 200\,keV, i.e., at and above the spectral cutoff. \textsl{INTEGRAL}'s IBIS and SPI are unique in that that can be used as polarimeters in this energy range, even though the measurements are challenging and require long observations.

The first measurement of hard X-ray/soft $\gamma$-ray polarization of a BH-XRB was presented by \citet{Laurent_2011a} using  \textsl{INTEGRAL}/IBIS: they show that the 20--2000\,keV spectrum of Cyg X-1 can be decomposed into a cutoff power law in the 20--400\,keV range and a hard tail above 400\,keV. The emission above 400\,keV is significantly polarized with a polarization fraction of $67 \pm 30$\%; below 400\,keV it is either weakly polarized or not polarized at all. The high degree of polarization is a sign of the jet origin of the hard tail. The result, including the separation of the two components with different polarization properties, was independently confirmed by \citet{Jourdain_2012a} using \textsl{INTEGRAL}/SPI. The first results were refined in the state-resolved polarization analysis of \citet{Rodriguez_2015a}, who used data until and including the year 2012. They show that the emission in the hard state is polarized, in agreement with \citet{Laurent_2011a} and \citet{Jourdain_2012a}, whose data were predominantly in the hard state \citep{Grinberg_2013a}. However, not enough observations could be collected in the intermediate state to allow a polarization analysis and only high upper limits on polarization can be obtained in the soft state. A detailed discussion of the most recent analysis update
is given elsewhere in this volume \citep{Cangemi_2019a}.

Bright, persistent, and often found in the hard state, Cyg X-1 is uniquely suited for polarization studies.
For other BH-XRBs, the detection of polarization is challenging as they lack brightness and are transient. Some first results for polarization of V404~Cyg have, however, been presented by \citet{Laurent_2016a}.

\section{Summary}

\textsl{INTEGRAL} has made not only major but also unique contributions to the field of BH-XRBs. Its high energy coverage remains crucial to understanding outburst of both known, new to \textsl{INTEGRAL}, and new sources.


\begin{acknowledgements} VG thanks the SOC and LOC of the 
 \textsl{INTEGRAL}2019 conference, especially C.~Ferrigno, E.~Bozzo and M.~Logossou, for the inspiring and smoothly organized meeting. VG is supported through the Margarete von Wrangell fellowship by the ESF and the Ministry of Science, Research and the Arts Baden-W\"urttemberg.
\end{acknowledgements}


\begin{thebibliography}{47}
\expandafter\ifx\csname natexlab\endcsname\relax\def\natexlab#1{#1}\fi

\bibitem[{{Cabanac} {et~al.}(2011){Cabanac}, {Roques}, \&
  {Jourdain}}]{Cabanac_2011a}
{Cabanac}, C., {Roques}, J.-P., \& {Jourdain}, E. 2011, \apj, 739, 58

\bibitem[{{Cangemi} {et~al.}(2019{\natexlab{a}}){Cangemi}, {Belloni}, \&
  {Rodriguez}}]{Cangemi_2019b}
{Cangemi}, F., {Belloni}, T., \& {Rodriguez}, J. 2019{\natexlab{a}}, ATel,
  12471

\bibitem[{{Cangemi} {et~al.}(2019{\natexlab{b}}){Cangemi}, {Beuchert},
  {Siegert}, {Grinberg}, {Wilms}, {Rodriguez}, {Kreykenbohm}, {Laurent}, \&
  {Pottschmidt}}]{Cangemi_2019a}
{Cangemi}, F., {Beuchert}, T., {Siegert}, T., {et~al.} 2019{\natexlab{b}},
  arXiv e-prints, arXiv:1904.09112

\bibitem[{{Del Santo}(2012)}]{Del_Santo_2012a}
{Del Santo}, M. 2012, in Journal of Physics Conference Series, Vol. 354,
  Journal of Physics Conference Series, 012003

\bibitem[{{Del Santo} {et~al.}(2016){Del Santo}, {Belloni}, {Tomsick},
  {Sbarufatti}, {Cadolle Bel}, {Casella}, {Castro-Tirado}, {Corbel},
  {Grinberg}, {Homan}, {Kalemci}, {Motta}, {Mu{\~n}oz-Darias}, {Pottschmidt},
  {Rodriguez}, \& {Wilms}}]{Del_Santo_2016a}
{Del Santo}, M., {Belloni}, T.~M., {Tomsick}, J.~A., {et~al.} 2016, \mnras,
  456, 3585

\bibitem[{{Del Santo} {et~al.}(2013){Del Santo}, {Malzac}, {Belmont},
  {Bouchet}, \& {De Cesare}}]{Del_Santo_2013a}
{Del Santo}, M., {Malzac}, J., {Belmont}, R.,  {et~al.}
  2013, \mnras, 430, 209

\bibitem[{{Dove} {et~al.}(1997){Dove}, {Wilms}, {Maisack}, \&
  {Begelman}}]{Dove_1997a}
{Dove}, J.~B., {Wilms}, J., {Maisack}, M., \& {Begelman}, M.~C. 1997, \apj,
  487, 759

\bibitem[{{Droulans} {et~al.}(2010){Droulans}, {Belmont}, {Malzac}, \&
  {Jourdain}}]{Droulans_2010a}
{Droulans}, R., {Belmont}, R., {Malzac}, J., \& {Jourdain}, E. 2010, \apj, 717,
  1022

\bibitem[{{Fender} {et~al.}(2004){Fender}, {Belloni}, \&
  {Gallo}}]{Fender_2004a}
{Fender}, R.~P., {Belloni}, T.~M., \& {Gallo}, E. 2004, \mnras, 355, 1105

\bibitem[{{Garc{\'{\i}}a} {et~al.}(2015){Garc{\'{\i}}a}, {Dauser}, {Steiner},
  {McClintock}, {Keck}, \& {Wilms}}]{Garcia_2015a}
{Garc{\'{\i}}a}, J.~A., {Dauser}, T., {Steiner}, J.~F., {et~al.} 2015, \apjl,
  808, L37

\bibitem[{{Garc{\'{\i}}a} {et~al.}(2016){Garc{\'{\i}}a}, {Grinberg}, {Steiner},
  {McClintock}, {Pottschmidt}, \& {Rothschild}}]{Garcia_2016a}
{Garc{\'{\i}}a}, J.~A., {Grinberg}, V., {Steiner}, J.~F., {et~al.} 2016, \apj,
  819, 76

\bibitem[{{Garc{\'{\i}}a} {et~al.}(2014){Garc{\'{\i}}a}, {McClintock},
  {Steiner}, {Remillard}, \& {Grinberg}}]{Garcia_2014a}
{Garc{\'{\i}}a}, J.~A., {McClintock}, J.~E., {Steiner}, {et~al.} 2014, \apj, 794, 73

\bibitem[{{Grinberg} {et~al.}(2011){Grinberg}, {Kreykenbohm}, {F{\"u}rst},
  {Wilms}, {Pottschmidt}, {Cadolle Bel}, {Rodriguez}, {Marcu}, {Suchy},
  {Markowitz}, \& {Nowak}}]{Grinberg_2011a}
{Grinberg}, V., {Kreykenbohm}, I., {F{\"u}rst}, F., {et~al.} 2011, Acta
  Polytechnica, 51, 33

\bibitem[{{Grinberg} {et~al.}(2013){Grinberg}, {Hell}, {Pottschmidt},
  {B{\"o}ck}, {Nowak}, {Rodriguez}, {Bodaghee}, {Cadolle Bel}, {Case}, {Hanke},
  {K{\"u}hnel}, {Markoff}, {Pooley}, {Rothschild}, {Tomsick}, {Wilson-Hodge},
  \& {Wilms}}]{Grinberg_2013a}
{Grinberg}, V., {Hell}, N., {Pottschmidt}, K., {et~al.} 2013, \aap, 554, A88


\bibitem[{{Grinberg} {et~al.}(2014){Grinberg}, {Pottschmidt}, {B{\"o}ck},
  {Schmid}, {Nowak}, {Uttley}, {Tomsick}, {Rodriguez}, {Hell}, {Markowitz},
  {Bodaghee}, {Cadolle Bel}, {Rothschild}, \& {Wilms}}]{Grinberg_2014a}
{Grinberg}, V., {Pottschmidt}, K., {B{\"o}ck}, M., {et~al.} 2014, \aap, 565, A1

\bibitem[{{Haardt} \& {Maraschi}(1991)}]{Haardt_1991a}
{Haardt}, F. \& {Maraschi}, L. 1991, \apjl, 380, L51

\bibitem[{{Homan} \& {Belloni}(2005)}]{Homan_2005a}
{Homan}, J. \& {Belloni}, T. 2005, \apss, 300, 107

\bibitem[{{Huppenkothen} {et~al.}(2017){Huppenkothen}, {Younes}, {Ingram},
  {Kouveliotou}, {G{\"o}{\u{g}}{\"u}{\c{s}}}, {Bachetti},
  {S{\'a}nchez-Fern{\'a}ndez}, {Chenevez}, {Motta}, {van der Klis}, {Granot},
  {Gehrels}, {Kuulkers}, {Tomsick}, \& {Walton}}]{Huppenkothen_2017a}
{Huppenkothen}, D., {Younes}, G., {Ingram}, A., {et~al.} 2017, \apj, 834, 90

\bibitem[{{Hynes} {et~al.}(2019){Hynes}, {Robinson}, {Terndrup}, {Gand hi},
  {Froning}, {Starrfield}, {Dhillon}, \& {Marsh}}]{Hynes_2019a}
{Hynes}, R.~I., {Robinson}, E.~L., {Terndrup}, D.~M., {et~al.} 2019, MNRAS
  accepted, arXiv:1905.00949

\bibitem[{{Joinet} {et~al.}(2007){Joinet}, {Jourdain}, {Malzac}, {Roques},
  {Corbel}, {Rodriguez}, \& {Kalemci}}]{Joinet_2007a}
{Joinet}, A., {Jourdain}, E., {Malzac}, J., {et~al.} 2007, \apj, 657, 400

\bibitem[{{Jourdain} {et~al.}(2012){Jourdain}, {Roques}, {Chauvin}, \&
  {Clark}}]{Jourdain_2012a}
{Jourdain}, E., {Roques}, J.~P., {Chauvin}, M., \& {Clark}, D.~J. 2012, \apj,
  761, 27

\bibitem[{{Jourdain} {et~al.}(2017){Jourdain}, {Roques}, \&
  {Rodi}}]{Jourdain_2017a}
{Jourdain}, E., {Roques}, J.-P., \& {Rodi}, J. 2017, \apj, 834, 130

\bibitem[{{Kalemci} {et~al.}(2014){Kalemci}, {Arabac{\i}}, {G{\"u}ver},
  {Russell}, {Tomsick}, {Wilms}, {Weidenspointner}, {Kuulkers}, {Falanga},
  {Din{\c c}er}, {Drave}, {Belloni}, {Coriat}, {Lewis}, \&
  {Mu{\~n}oz-Darias}}]{Kalemci_2014a}
{Kalemci}, E., {Arabac{\i}}, M.~{\"O}., {G{\"u}ver}, T., {et~al.} 2014, \mnras,
  445, 1288

\bibitem[{{Kuulkers} {et~al.}(2018){Kuulkers}, {Bozzo}, {Ferrigno}, {Belloni},
  \& {Sanchez-Fernandez}}]{Kuulkers_2018a}
{Kuulkers}, E., {Bozzo}, E., {Ferrigno}, C.,  {et~al.} 2018, ATel, 11490

\bibitem[{{Laurent} {et~al.}(2016){Laurent}, {Gouiffes}, {Rodriguez}, \&
  {Chambouleyron}}]{Laurent_2016a}
{Laurent}, P., {Gouiffes}, C., {Rodriguez}, J., \& {Chambouleyron}, V. 2016, in
  Proceedings of the 11th INTEGRAL Conference, 10-14 October 2016 Amsterdam,
  NL, 22

\bibitem[{{Laurent} {et~al.}(2011){Laurent}, {Rodriguez}, {Wilms}, {Cadolle
  Bel}, {Pottschmidt}, \& {Grinberg}}]{Laurent_2011a}
{Laurent}, P., {Rodriguez}, J., {Wilms}, J., {et~al.} 2011, Science, 332, 438

\bibitem[{{Maitra} {et~al.}(2017){Maitra}, {Scarpaci}, {Grinberg}, {Reynolds},
  {Markoff}, {Maccarone}, \& {Hynes}}]{Maitra_2017a}
{Maitra}, D., {Scarpaci}, J.~F., {Grinberg}, V., {et~al.} 2017, \apj, 851, 148

\bibitem[{{Markoff} {et~al.}(2005){Markoff}, {Nowak}, \&
  {Wilms}}]{Markoff_2005a}
{Markoff}, S., {Nowak}, M.~A., \& {Wilms}, J. 2005, \apj, 635, 1203

\bibitem[{{Matt} {et~al.}(1992){Matt}, {Perola}, {Piro}, \&
  {Stella}}]{Matt_1992a}
{Matt}, G., {Perola}, G.~C., {Piro}, L., \& {Stella}, L. 1992, \aap, 257, 63

\bibitem[{{McConnell} {et~al.}(2000){McConnell}, {Ryan}, {Collmar},
  {Sch{\"o}nfelder}, {Steinle}, {Strong}, {Bloemen}, {Hermsen}, {Kuiper},
  {Bennett}, {Phlips}, \& {Ling}}]{McConnell_2000a}
{McConnell}, M.~L., {Ryan}, J.~M., {Collmar}, W., {et~al.} 2000, \apj, 543, 928

\bibitem[{{McConnell} {et~al.}(2002){McConnell}, {Zdziarski}, {Bennett},
  {Bloemen}, {Collmar}, {Hermsen}, {Kuiper}, {Paciesas}, {Phlips}, {Poutanen},
  {Ryan}, {Sch{\"o}nfelder}, {Steinle}, \& {Strong}}]{McConnell_2002a}
{McConnell}, M.~L., {Zdziarski}, A.~A., {Bennett}, K., {et~al.} 2002, \apj,
  572, 984

\bibitem[{{Motta} {et~al.}(2017){Motta}, {Kajava}, {S{\'a}nchez-Fern{\'a}ndez},
  {Giustini}, \& {Kuulkers}}]{Motta_2017a}
{Motta}, S.~E., {Kajava}, J.~J.~E., {S{\'a}nchez-Fern{\'a}ndez}, C.,
  {Giustini}, M., \& {Kuulkers}, E. 2017, \mnras, 468, 981

\bibitem[{{Mu{\~n}oz-Darias} {et~al.}(2016){Mu{\~n}oz-Darias}, {Casares}, {Mata
  S{\'a}nchez}, {Fender}, {Armas Padilla}, {Linares}, {Ponti}, {Charles},
  {Mooley}, \& {Rodriguez}}]{Munoz-Darias_2016a}
{Mu{\~n}oz-Darias}, T., {Casares}, J., {Mata S{\'a}nchez}, D., {et~al.} 2016,
  \nat, 534, 75

\bibitem[{{Nandra} {et~al.}(2013){Nandra}, {Barret}, {Barcons}, {Fabian}, {den
  Herder}, {Piro}, {Watson}, {Adami}, {Aird}, {Afonso}, \&
  et~al.}]{Nandra_2013a}
{Nandra}, K., {Barret}, D., {Barcons}, X., {et~al.} 2013, arXiv e-prints,
  arXiv:1306.2307

\bibitem[{{Natalucci} {et~al.}(2015){Natalucci}, {Fiocchi}, {Bazzano},
  {Ubertini}, {Roques}, \& {Jourdain}}]{Natalucci_2015a}
{Natalucci}, L., {Fiocchi}, M., {Bazzano}, A., {et~al.} 2015, \apj, 813, L21

\bibitem[{{Nowak} {et~al.}(2011){Nowak}, {Hanke}, {Trowbridge}, {Markoff},
  {Wilms}, {Pottschmidt}, {Coppi}, {Maitra}, {Davis}, \&
  {Tramper}}]{Nowak_2011a}
{Nowak}, M.~A., {Hanke}, M., {Trowbridge}, S.~N., {et~al.} 2011, \apj, 728, 13

\bibitem[{{Parmar} {et~al.}(2003){Parmar}, {Kuulkers}, {Oosterbroek}, {Barr},
  {Much}, {Orr}, {Williams}, \& {Winkler}}]{Parmar_2003a}
{Parmar}, A.~N., {Kuulkers}, E., {Oosterbroek}, T., {et~al.} 2003, \aap, 411,
  L421

\bibitem[{{Rodriguez} {et~al.}(2015{\natexlab{a}}){Rodriguez}, {Cadolle Bel},
  {Alfonso-Garz{\'o}n}, {Siegert}, {Zhang}, {Grinberg}, {Savchenko}, {Tomsick},
  {Chenevez}, {Clavel}, {Corbel}, {Diehl}, {Domingo}, {Gouiff{\`e}s},
  {Greiner}, {Krause}, {Laurent}, {Loh}, {Markoff}, {Mas-Hesse},
  {Miller-Jones}, {Russell}, \& {Wilms}}]{Rodriguez_2015b}
{Rodriguez}, J., {Cadolle Bel}, M., {Alfonso-Garz{\'o}n}, J., {et~al.}
  2015{\natexlab{a}}, \aap, 581, L9

\bibitem[{{Rodriguez} {et~al.}(2015{\natexlab{b}}){Rodriguez}, {Grinberg},
  {Laurent}, {Cadolle Bel}, {Pottschmidt}, {Pooley}, {Bodaghee}, {Wilms}, \&
  {Gouiff{\`e}s}}]{Rodriguez_2015a}
{Rodriguez}, J., {Grinberg}, V., {Laurent}, P., {et~al.} 2015{\natexlab{b}},
  \apj, 807, 17

\bibitem[{{Roques} \& {Jourdain}(2016)}]{Roques_2016a}
{Roques}, J.~P. \& {Jourdain}, E. 2016, arXiv e-prints, arXiv:1601.05289

\bibitem[{{Roques} \& {Jourdain}(2019)}]{Roques_2019a}
{Roques}, J.-P. \& {Jourdain}, E. 2019, \apj, 870, 92

\bibitem[{{S{\'a}nchez-Fern{\'a}ndez}
  {et~al.}(2017){S{\'a}nchez-Fern{\'a}ndez}, {Kajava}, {Motta}, \&
  {Kuulkers}}]{Sanchez-Fernandez_2017a}
{S{\'a}nchez-Fern{\'a}ndez}, C., {Kajava}, J.~J.~E., {Motta}, S.~E., \&
  {Kuulkers}, E. 2017, \aap, 602, A40

\bibitem[{{Siegert} {et~al.}(2016){Siegert}, {Diehl}, {Greiner}, {Krause},
  {Beloborodov}, {Bel}, {Guglielmetti}, {Rodriguez}, {Strong}, \&
  {Zhang}}]{Siegert_2016a}
{Siegert}, T., {Diehl}, R., {Greiner}, J., {et~al.} 2016, \nat, 531, 341

\bibitem[{{Walton} {et~al.}(2016){Walton}, {Tomsick}, {Madsen}, {Grinberg},
  {Barret}, {Boggs}, {Christensen}, {Clavel}, {Craig}, {Fabian}, {Fuerst},
  {Hailey}, {Harrison}, {Miller}, {Parker}, {Rahoui}, {Stern}, {Tao}, {Wilms},
  \& {Zhang}}]{Walton_2016a}
{Walton}, D.~J., {Tomsick}, J.~A., {Madsen}, K.~K., {et~al.} 2016, \apj, 826,
  87

\bibitem[{{Weisskopf} {et~al.}(2016){Weisskopf}, {Ramsey}, {O'Dell}, {Tennant},
  {Elsner}, {Soffitta}, {Bellazzini}, {Costa}, {Kolodziejczak}, {Kaspi},
  {Muleri}, {Marshall}, {Matt}, \& {Romani}}]{Weisskop_2016a}
{Weisskopf}, M.~C., {Ramsey}, B., {O'Dell}, S., {et~al.} 2016, in SPIE
  Conference Series, Vol. 9905, Space Telescopes and Instrumentation 2016:
  Ultraviolet to Gamma Ray, 990517

\bibitem[{{Zdziarski} {et~al.}(2012){Zdziarski}, {Lubi{\'n}ski}, \&
  {Sikora}}]{Zdziarski_2012a}
{Zdziarski}, A.~A., {Lubi{\'n}ski}, P., \& {Sikora}, M. 2012, \mnras, 423, 663

\bibitem[{{Zhang} {et~al.}(2016){Zhang}, {Feroci}, {Santangelo}, {Dong},
  {Feng}, {Lu}, {Nandra}, {Wang}, {Zhang}, {Bozzo}, {Brandt}, {De Rosa}, {Gou},
  {Hernanz}, {van der Klis}, {Li}, {Liu}, {Orleanski}, {Pareschi}, {Pohl},
  {Poutanen}, {Qu}, {Schanne}, {Stella}, {Uttley}, {Watts}, {Xu}, {Yu}, {in 't
  Zand}, {Zane}, {Alvarez}, {Amati}, {Baldini}, {Bambi}, {Basso},
  {Bhattacharyya S.}, {}, {Belloni}, {Bellutti}, {Bianchi}, {Brez}, {Bursa},
  {Burwitz}, {Budtz-J{\o}rgensen}, {Caiazzo}, {Campana}, {Cao}, {Casella},
  {Chen}, {Chen}, {Chen}, {Chen}, {Chen}, {Chen}, {Civitani}, {Coti Zelati},
  {Cui}, {Cui}, {Dai}, {Del Monte}, {de Martino}, {Di Cosimo}, {Diebold},
  {Dovciak}, {Donnarumma}, {Doroshenko}, {Esposito}, {Evangelista}, {Favre},
  {Friedrich}, {Fuschino}, {Galvez}, {Gao}, {Ge}, {Gevin}, {Goetz}, {Han},
  {Heyl}, {Horak}, {Hu}, {Huang}, {Huang}, {Hudec}, {Huppenkothen}, {Israel},
  {Ingram}, {Karas}, {Karelin}, {Jenke}, {Ji}, {Korpela}, {Kunneriath},
  {Labanti}, {Li}, {Li}, {Li}, {Liang}, {Limousin}, {Lin}, {Ling}, {Liu},
  {Liu}, {Liu}, {Lu}, {Lund}, {Lai}, {Luo}, {Luo}, {Ma}, {Mahmoodifar},
  {Marisaldi}, {Martindale}, {Meidinger}, {Men}, {Michalska}, {Mignani},
  {Minuti}, {Motta}, {Muleri}, {Neilsen}, {Orlandini}, {Pan}, {Patruno},
  {Perinati}, {Picciotto}, {Piemonte}, {Pinchera}, {Rachevski A.}, {Rapisarda},
  {Rea}, {Rossi}, {Rubini}, {Sala}, {Shu}, {Sgro}, {Shen}, {Soffitta}, {Song},
  {Spandre}, {Stratta}, {Strohmayer}, {Sun}, {Svoboda}, {Tagliaferri},
  {Tenzer}, {Hong}, {Taverna}, {Torok}, {Turolla}, {Vacchi}, {Wang}, {Walton},
  {Wang}, {Wang}, {Wang}, {Wang}, {Weng}, {Wilms}, {Winter}, {Wu}, {Wu},
  {Xiong}, {Xu}, {Xue}, {Yan}, {Yang}, {Yang}, {Yang}, {Yuan}, {Yuan}, {Yuan},
  {Zampa}, {Zampa}, {Zdziarski}, {Zhang}, {Zhang}, {Zhang}, {Zhang}, {Zhang},
  {Zhang}, {Zheng}, {Zhou}, \& {Zhou X.~L.}}]{Zhang_2016a}
{Zhang}, S.~N., {Feroci}, M., {Santangelo}, A., {et~al.} 2016, in SPIE
  Conference Series, Vol. 9905, Space Telescopes and Instrumentation 2016:
  Ultraviolet to Gamma Ray, 99051Q

\end{thebibliography}

\end{document}